# Electromagnetic Effects on Strongly Interacting QCD-Matter

Abdel Magied Abdel Aal DIAB[1,2*], Abdel Nasser TAWFIK[1,2†] and M. T. HUSSEIN[3‡]

[1]*Egyptian Center for Theoretical Physics (ECTP), Modern University for Technology and Information (MTI), 11571 Cairo, Egypt*
[2]*World Laboratory for Cosmology And Particle Physics (WLCAPP), Cairo, Egypt*
[3]*Physics Department, Faculty of Science, Cairo University, 12613 Giza, Egypt*

In order to study the temperature dependence of the quark-hadron phase structure and the QCD equation of state in vanishing and finite magnetic field, the SU(3) Polyakov linear-sigma model is utilized. In mean field approximation, the dependence of various magnetic properties such as magnetization, magnetic susceptibility and magnetic catalysis is analyzed in finite magnetic field. Furthermore, the influences of finite magnetic field on the temperature dependence of some transport properties (viscosity and conductivity) from Green-Kubo correlation are characterized.

**INTRODUCTION**
Due to oppositely directed off-central relativistic collision of two heavy ions, a huge magnetic field ($eB$) can be created in heavy-ion collision (HIC) experiments. At SPS, RHIC and LHC energies, the strength of this field ranges from 0.1 to 1 and 15 $m_\pi^2$, respectively [1, 2]. In fact, such field is much localized and short lived and thus it is assumed to have almost no effect on the detectors and their external magnet but remarkable influences on the strongly interacting quantum chromodynamics (QCD) matter.

In the present work, we utilize the Polyakov linear-sigma model (PLSM) in analyzing the temperature dependence of the quark-hadron phase structure and the QCD equation of state (EoS) in vanishing and finite magnetic field. We also present estimation for the magnetization ($\mathcal{M}$), the magnetic susceptibility ($\chi_B$) and the magnetic catalysis of as functions of temperature and compare them to recent lattice QCD calculations. Last but not least, we introduce a direct estimation for various transport properties such as bulk ($\xi$) and shear ($\eta$) viscosities and electric ($\sigma_e$) and thermal ($\kappa$) conductivities from the Green-Kubo (GK) correlations in finite magnetic field.

The present paper is organized as follows. A short reminder to SU(3) PLSM in mean field approximation shall be presented in section II. Section III summarizes our calculations at finite temperature and finite magnetic field. Section IV is devoted to the conclusions.

**THE APPROACH**
PLSM is a successful approach for the quark-hadron phase transition(s), the QCD thermodynamic quantities, and the transport properties. Details about the model can be found in Ref. [3]. In the mean field approximation, the exchange of energy between particle and antiparticle at temperature (T) and baryon chemical potential (μ) is described by the grand canonical partition function ($\mathcal{Z}$). The free energy in finite volume (V) is given as $\mathcal{F} = -\text{T}.\log[\mathcal{Z}]/V$ or

[*] a.diab@eng.mti.edu.eg
[†] a.tawfik@eng.mti.edu.eg
[‡] tarek@sci.cu.edu.eg

$$\mathcal{F} = U(\sigma_l, \sigma_s) + \mathcal{U}(\varphi, \varphi^*, T) + \Omega_{\bar{q}q}(T, \mu, B) + \delta_{0,eB}\, \Omega_{\bar{q}q}(T, \mu),$$

where $U(\sigma_l, \sigma_s)$ is the pure mesonic potential [3] and the second term represents the Polyakov loop potential. In the present work, we use polynomial expansion for $\mathcal{U}(\varphi, \varphi^*, T)$ in $\varphi$ and $\varphi^*$ [4]. The quarks and anti-quark contributions to the medium potential $\Omega_{\bar{q}q}$ can be divided into two regimes:

- In vanishing magnetic field ($eB = 0$),

$$\Omega_{\bar{q}q}(T, \mu) = -2T \sum_{f=l,s} \int \frac{d^3 p}{(2\pi)^3} \{\ln g_f^+ + \ln g_f^-\},$$

where $g_f^\pm$ is the quark and antiquark contribution, respectively.

- In finite magnetic field ($eB \neq 0$), the concepts of Landau quantization and magnetic catalysis can be implemented. The magnetic field is assumed to be oriented along $z-$ direction. Firstly, the finite magnetic field modifies the quarks dispersion relation,

$$E_{B,f} = \left[p_z^2 + m_f^2 + |q_f|(2n + 1 - \sigma)B\right]^{1/2},$$

where $q_f$ is the electric charge of $f^{\text{th}}$ quark-flavor, $n$ is the quantized landau number and $\sigma$ is related to the spin quantum number S; $\sigma = \pm S/2$. Secondly, at $eB \neq 0$ the potential becomes

$$\Omega_{\bar{q}q}(T, \mu, B) = -2 \sum_{f=l,s} \frac{|q_f|BT}{(2\pi)^2} \sum_{\nu=0}^{\infty} (2 - \delta_{0\nu}) \int d p_z \{\ln g_f^+ + \ln g_f^-\},$$

where the $(2 - \delta_{0\nu})$ term stands for degenerate Landau levels and the $(2n + 1 - \sigma)$ term is replaced by the number of Landau levels $\nu$.

**RESULTS**

At $eB \neq 0$, we introduce various properties of the QCD matter such as the magnetization ($\mathcal{M}$), the magnetic susceptibility ($\chi_B$), the magnetic catalysis and the EoS as functions of temperature and confront them to recent lattice QCD calculations. We also present numerical estimations for bulk and shear viscosity normalized to the thermal entropy (s) and the dimensionless electric and thermal conductivities, $\xi/s$, $\eta/s$, $\sigma_e/T$ and $\kappa/T^2$, respectively.

As assumed, the influence of the magnetic field is limited to $z-$direction, thus $B = B\, \hat{e}_z$, and the free energy density $f = \mathcal{F}/V$ gets modifications as follows [5],

$$f = \epsilon^{\text{tot}} - \epsilon^{\text{field}} - Ts = \epsilon^{\text{tot}} - Ts - eB\, \mathcal{M},$$

where $\epsilon^{\text{tot}} = \epsilon + \epsilon^{\text{field}}$ is the total energy density including the energy density ($\epsilon$) of the system and $\epsilon^{\text{field}} = eB\mathcal{M}$ stemming from the influence of the magnetic field. The magnetization ($\mathcal{M}$) and magnetic susceptibility ($\chi_B$) can be derived as,

$$\mathcal{M} = -\frac{1}{V} \frac{\partial \mathcal{F}}{\partial (eB)}, \qquad \chi_B = -\frac{1}{V} \frac{\partial^2 \mathcal{F}}{\partial (eB)^2}\bigg|_{eB=0}.$$

Fig. 1 depicts the temperature dependence of the magnetization [left-hand panel (a)] in GeV$^2$ at $eB \neq 0$ and the magnetic susceptibility($\chi_B$) [right-hand panel (b)] at $eB = 0$. The PLSM magnetization is confronted to recent lattice QCD calculation [5]. It is obvious that the magnetization ($\mathcal{M} > 0$) increases with the temperature. This dependence reflects that the thermal QCD-medium is paramagnetic around and above the critical temperature. We also observe an evidence for weak diamagnetism at low temperatures; the magnetization decreases with the temperature, inside box.

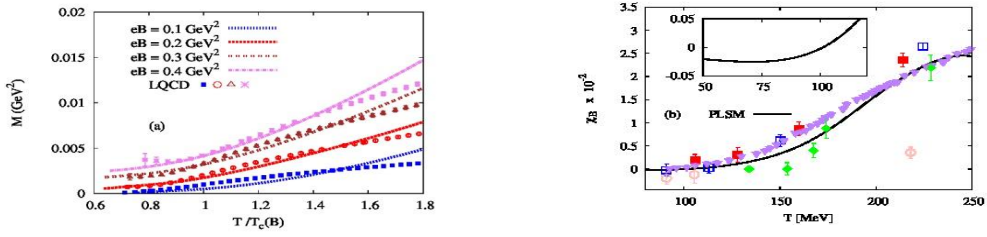

**Fig. 1:** Left-hand panel (a) presents the temperature dependence of the magnetization ($\mathcal{M}$) at $eB \neq 0$. Right-hand panel (b) shows the magnetic susceptibility($\chi_B$) as a function of temperature at $eB = 0$. The PLSM results are compared with recent lattice QCD simulations (symbols) [5].

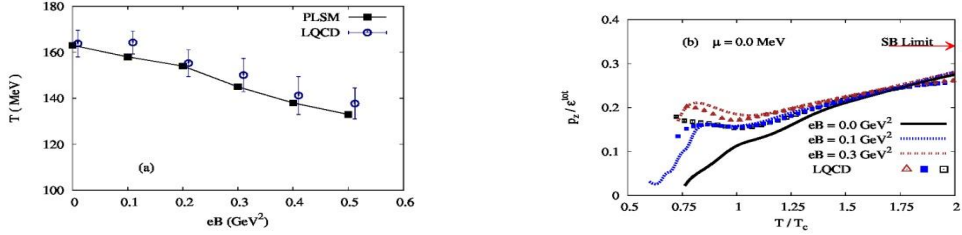

**Fig. 2:** Left-hand panel (a) depicts the dependence of the critical temperature on $B$. The ratio $p_z/\epsilon^{tot}$ is given as function of $T/T_c$ at $eB \neq 0$. The PLSM results are compared with recent lattice QCD simulations (symbols) [5].

Left-hand panel of Fig. 2 (a) shows the variation of the critical temperature with increasing magnetic field. The PLSM critical temperature is estimated from the intersection point (temperature) between the subtracted chiral condensate and the deconfinement phase transition. It is apparent that the critical temperature decreases with raising magnetic field. This phenomenon is known as inverse magnetic catalysis. The speed of sound squared $c_s^2$ plays an essential role in the estimation of EoS $p_z(\epsilon^{tot})$. Right-hand panel of Fig. 2 (b) presents the PLSM calculations for $p_z/\epsilon^{tot}$ at $eB = 0$ (solid curve), $eB = 0.1$ (dotted curve) and $eB = 0.3$ GeV$^2$ (double-dotted curve). These results are compared with recent lattice QCD [5] (open square), (close square), and (open triangle), respectively. A good agreement is achieved in comparing the PLSM results with the lattice QCD. It seems that the agreement can be improved with increasing the magnetic field strength. For further PLSM results in presence of the magnetic field the readers are advised to consult Ref. [6].

The response of QCD matter to the presence of finite magnetic field can be described by the transport coefficients, as well. From the Green-Kubo (GK) correlations, the bulk and shear viscosities ($\xi$ and $\eta$) are related to the correlation functions of the trace of the energy-momentum tensor [7].

Fig. 3 depicts the ratios $\xi/s$ and $\eta/s$ as functions of temperature in finite magnetic field. The solid curve represents the results at $eB = 0.0$, while $eB = 0.2$ and $eB = 0.4$ GeV$^2$ are given as dotted, dot-dashed curves, receptively. The Kovtun, Son and Starinets (KSS) limit is represented by the dashed line. At low temperatures, $\xi/s$ and $\eta/s$ becomes very large. A peak positioned at $T_c$ starts to appear. At extremely high temperatures, the coupling seems to become weak and the hadrons liberate into quarks and gluons. Therefore, the temperature dependent of both quantities becomes very weak.

The response of QGP-matter for finite electric field can be determined by the electric conductivity $\sigma_e$, which is related to the flow of the charges. Left-hand panel of Fig. 4 (a) shows $\sigma_e/T$ as a function of temperature and compares this with recent lattice QCD simulations [9-11] and different QCD-like models such as NJL and DQPM [8].

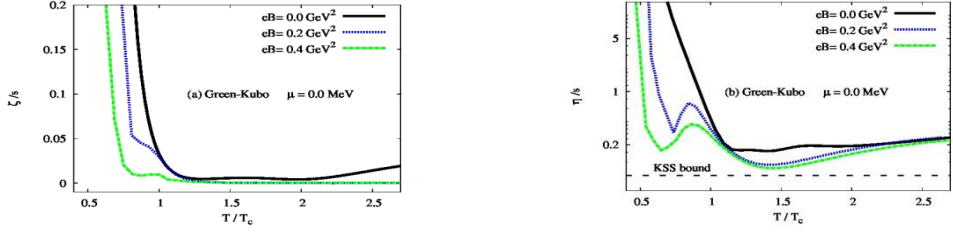

**Fig. 3:** Left-hand panel (a) depicts $\xi/s$ as a function of temperature in finite magnetic field $eB = 0.0$ (solid curve), $eB = 0.2$ and $eB = 0.4$ GeV$^2$ (dotted dash curve). Right-hand panel (b) shows the same as left-hand panel (a) but for $\eta/s$.

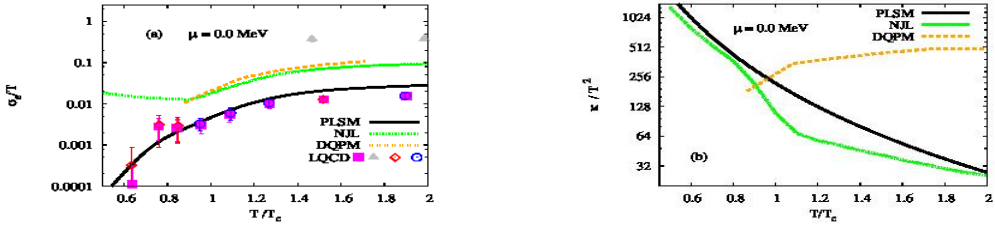

**Fig. 4:** Left-hand panel (a) shows $\sigma_e/T$ as a function of temperature. PLSM calculations (solid curve) are confronted to recent lattice QCD simulations [9-11] and also compared with NJL and DQPM models [8]. Right-hand panel (b) presents $\kappa/T^2$ as a function of temperature. PLSM calculation (solid curve) is compared with NJL and DQPM models [8].

The heat conductivity $\kappa$ is related to the heat flow in such relativistic fluid. From relativistic Navier-Stokes ansatz, the heat flow is proportional to the gradient of thermal potential. Right-hand panel (b) shows $\kappa/T^2$ as a function of temperature. The results are compared with NJL and DQPM models [8]. There are no lattice QCD calculations to compare with it.

**CONCLUSION**
The PLSM seems to be able to reproduce recent lattice QCD calculations for the magnetization, the magnetic susceptibility, the magnetic catalysis and the transport properties (viscosity and conductivity) in vanishing and finite magnetic field. We conclude that the QCD matter possesses paramagnetic property above the critical temperature but diamagnetic one at low temperatures. The PLSM confirms that inverse magnetic catalysis.

**REFERENCES**
1. V. Skokov, A. Y. Illarionov, and V. Toneev, Int. J. Mod. Phys. A **24**, 5925 (2009).
2. A. Bzdak and V. Skokov, Phys. Lett. B **710**, 174 (2012).
3. A. Tawfik and A. Diab, Phys. Rev. C **91**, 015204 (2015).
4. C. Ratti, M. A. Thaler, and W. Weise, Phys. Rev. D **73**, 014019 (2005).
5. G.S. Bali, F. Bruckmann, G. Endrdi, S.D. Katz, and A. Schaefer, JHEP **1408**, 177 (2014).
6. A. Tawfik and A. Diab, and M.T. Hussein, arXiv:1604.08174 [hep-lat].
7. D. Fernandez-Fraile and A. Gomez Nicola, Eur. Phys. J. C **62**, 37 (2009).
8. R. Marty, *et. al.*, Phys. Rev. C **88**, 045204 (2013).
9. A. Amato, *et. al.,* Phys. Rev. Lett. **111**, 172001 (2013).
10. H.-T. Ding, *et. al.,* Phys. Rev. D **83**, 034504 (2011).
11. B. B. Brandt, *et. al.,* JHEP **1303**,100 (2013).